\documentstyle[aps]{revtex}

\begin{document}

\wideabs{


\title{Onset of Delocalization in Quasi-1D Waveguides\ \\
       with Correlated Surface Disorder}

\author{F.~M.~Izrailev}

\address{Instituto de F\'{\i}sica, Universidad Aut\'{o}noma de Puebla, \\
         Apartado Postal J-48, Puebla, Pue., 72570, M\'{e}xico}

\author{N.~M.~Makarov}

\address{Instituto de Ciencias, Universidad Aut\'{o}noma
         de Puebla, \\ Priv. 17 Norte No 3417, Col. San Miguel
         Hueyotlipan, Puebla, Pue., 72050, M\'{e}xico}

\date{\today}
\maketitle

\begin{abstract}
We present first analytical results on transport properties of
many-mode waveguides with rough surfaces having long-range
correlations. We show that propagation of waves through such
waveguides reveals a quite unexpected phenomena of a complete
transparency for a subset of propagating modes. These modes do not
interact with each other and effectively can be described by the
theory of 1D transport with correlated disorder. We also found
that with a proper choice of model parameters one can arrange a
perfect transparency of waveguides inside a given window of energy
of incoming waves. The results may be important in view of
experimental realizations of a selective transport in application
to both waveguides and electron/optic nanodevices.
\end{abstract}


} 


During last few years there is a burst of interest to solid state
models with correlated disorder (see, e.g., \cite{nature} and
references therein). This fact is mainly due to the possibility to
observe an anomalous transport in 1D models with random correlated
potentials. In particular, it was shown
\cite{our} that specific long-range
correlations in potentials give rise to a complete transparency of
electron waves for given energy windows. Experimental realization
of such potentials for single-mode waveguides with delta-like
scatters \cite{KIKS00} has confirmed theoretical predictions.

The subject of wave propagation through surface-disordered
waveguides (see,
e.g.,\cite{BFb79,makarov,SMFY99,falko,garcia,MT98} and references
therein) is important both from the theoretical viewpoint and for
experimental applications such as optic fibers, remote sensing,
radio wave propagation, shallow water waves, etc. On the other
hand, it describes electronic transport in mesoscopic conducting
channels as well. So far, main results in this field are obtained
for random surfaces with {\it fast-decaying} correlations along a
scattering surface. It is now of great importance to understand
main properties of multiple scattering from rough surfaces with
{\it long-range} correlations.

In this contribution we theoretically analyze transport properties
of quasi-1D waveguides with rough surfaces. Our main interest is
in exploring the possibility of constructing such surface profiles
that result in frequency windows of a perfect propagation of
waves.


In what follows we consider a plane waveguide (or conducting wire)
of average width $d$ and length $L$, stretched along the $x$-axis.
One (lower) surface of the waveguide is a rough profile
$z=\xi(x)$, and the other (upper) surface, for simplicity, is
taken to be flat, $z=d$. The function $\xi(x)$ describes the
boundary roughness and is assumed to be statistically homogeneous
and isotropic, with standard statistical properties,
\begin{equation}
\langle\xi(x)\rangle=0, \qquad
\langle\xi(x)\xi(x')\rangle=\sigma^2{\cal W}(|x-x'|).
\label{ksi}
\end{equation}
Here the angular brackets stand for a statistical average over
different realizations of a profile $\xi(x)$. The binary
correlator ${\cal W}(|x-x'|)$ is assumed to decrease on a
characteristic scale $k_0^{-1}$, and is normalized to its maximal
value, ${\cal W}(0)=1$.

Both the {\it amplitude} and {\it gradient} roughness of the
corrugated surface is taken to be small, $\sigma \ll d$ and $k_0
\sigma \ll 1$. These limitations are common in surface scattering
theories that are based on an appropriate {\it perturbative
approach}, see for example, \cite{BFb79}. We should also add that
zero (Dirichlet) boundary conditions are applied at both surfaces
of the waveguide.


Let us first consider a single-mode waveguide in which only one
normal mode ({\it channel}) has real lengthwise wave number
$k_1=\sqrt{k^2-(\pi /d)^2}$ (for details see \cite{IM01}). It is
known
\cite{MakYur89} that in this case the problem of surface
scattering is reduced to a purely 1D model with the potential
determined by the random profile $\xi(x)$. The corresponding wave
equation takes the form,
\begin{equation}\label{1DWeq}
\left(\frac{d^2}{dx^2}+k_1^2\right)\Psi(x)-
\frac{2}{\pi}\,\left(\frac{\pi}{d}\right)^3
\xi(x)\,\Psi(x)=0.
\end{equation}
Since one channel case corresponds to inequality $1
< kd/\pi < 2$, the value of $k_1$ is restricted by the relation
$0<k_1d/\pi<\sqrt{3}$. Note also that the weak scattering
assumption $\sigma\ll d$ leads to the condition, $k_1 \sigma \ll
1$.

The above equation (\ref{1DWeq}) demonstrates an exact
correspondence between the surface scattering in one-mode
waveguides, and bulk scattering for electrons in 1D models of the
Anderson (or Kronig-Penney) type. Specifically, the wave number
$k$ is equal to $\omega/c$ for a classical wave of frequency
$\omega$, and to the Fermi wave number for electrons.

In order to solve the wave equation (\ref{1DWeq}) and to find the
averaged {\it transmittance}, or, the same, the dimensionless {\it
conductance} (in units of $e^2/\pi\hbar$), one can use methods
developed in the transport theory of one-dimensional disordered
systems, such as the perturbative diagrammatic method of
Berezinski \cite{Ber73}, invariant imbedding method
\cite{BelWing75}, or the two-scale approach
\cite{MT98}. All the methods allow to take adequately
into account the effects of coherent multiple scattering of waves
(electrons) from the corrugated surface.

Main theoretical result is that the average transmittance $\langle
T\rangle$ is described by a universal expression which depends on
one parameter $\Lambda = L_{loc}/L$ only ({\it one-parameter
scaling}). Here the quantity $L_{loc}/4$ is the backscattering
length in a 1D disordered system. Its inverse value $L_{loc}^{-1}$
is equal to the Lyapunov exponent that can be found in the
transfer matrix approach, so that the quantity $L_{loc}$ is known
as the {\it localization length} \cite{LGP88}.

The expression for $\langle T\rangle$ is fully consistent with the
theory of 1D {\it Anderson localization}. Specifically, the
transmittance $\langle T\rangle$ exhibits the ballistic behavior
for large localization length, $L_{loc}/L\gg 1$,
\begin{equation}\label{1Dav-g-bal}
\langle T(L/L_{loc})\rangle\approx 1-4L/L_{loc},
\end{equation}
therefore, in this case the waveguide is practically transparent.
On the contrary, the transmittance is exponentially small when the
localization length is much less than the length of the waveguide,
$L_{loc}/L\ll 1$,
\begin{equation}\label{1Dav-g-loc}
\langle T(L/L_{loc})\rangle\approx\frac{\pi^{5/2}}{16}
\left(L/L_{loc}\right)^{-3/2}\exp\left(-L/L_{loc}\right).
\end{equation}
This implies strong wave localization. Since in this case the
transmittance is not a self-averaged quantity, for completeness of
presentation one should refer to the expression for its logarithm,
$\langle\ln T(L/L_{loc})\rangle =-4L/L_{loc}$.

Thus, in order to describe transport properties of a disordered
waveguide, one needs to know the  value of the localization length
$L_{loc}$. The latter is entirely determined by the type of
disorder. For the surface disorder, from the solution of the
equation (\ref{1DWeq}) one can obtain
\cite{MakYur89},
\begin{equation}\label{Lloc}
\frac{1}{L_{loc}}=\frac{\sigma^2}{\pi^2}\left(\frac{\pi}{d}\right)^6
\frac{W(2k_1)}{(2k_1)^2}.
\end{equation}
This formula contains the Fourier transform $W(k_x)$ of the binary
correlator ${\cal W}(|x-x'|)$, which is called the {\it roughness
power spectrum},
\begin{equation}
{\cal W}(|x-x'|)=\int_{-\infty}^{\infty}\frac{dk_x}{2\pi}
\exp\{ik_x(x-x')\}\,W(k_x).
\label{FR-W}
\end{equation}


One can see that the localization length and, correspondingly,
transport properties of a single-mode waveguide are completely
defined by the dependence $W(k_x)$. In particular, if $W(2k_1)$
abruptly vanishes within some interval of wave number $k_1$, then
the localization length $L_{loc}$ diverges and the waveguide of
any length will be fully transparent. Therefore, an interesting
problem arises: how to construct such (random) surface profiles
that result in a complete transparency of waveguides within a
predefined part of the allowed region for the wave vector $k_1$?
The answer to this question was done in Ref.\cite{IM01} where the
results obtained for discrete Anderson-like disordered models
\cite{our} have been extended to the surface scattering.

The recipe is as follows \cite{IM01}. Having a desirable
dependence for the Fourier transform $W(k_x)$, one obtains the
function $\beta(x)$,
\begin{equation}\label{beta-def}
\beta(x)=\int_{-\infty}^{\infty}\frac{dk_x}{2\pi}
\exp\left(ik_xx\right)\,W^{1/2}(k_x).
\end{equation}
Then, the surface profile can be constructed by the convolution of
the delta-correlated random process (white noise $Z(x)$) with the
function $\beta(x)$,
\begin{equation}\label{xi-beta}
\xi(x)=\sigma\,\int_{-\infty}^\infty\,dx'\,Z(x-x')\,\beta(x').
\end{equation}

The proposed method allows us to practically solve the inverse
scattering problem of a reconstruction of rough surfaces from
their power spectrum. We would like to emphasize that by a proper
choice of the binary correlator, the transition between localized
and delocalized wave/electron transport can occur abruptly at a
given point inside the allowed interval for the wave number $k_1$
(see details and examples in
\cite{IM01}). This means that the roughness power spectrum
$W(k_x)$ is discontinuous at this point. Consequently, the
correlator ${\cal W}(|x-x'|)$ of the corrugated surface $\xi(x)$
should be of {\it specific} form with {\it long-range
correlations} along the waveguide.

It is important to stress that surfaces with the above discussed
properties (when $W(k_x)$ vanishes in some regions of $k_x$) are
not exotic. They have been recently fabricated in experimental
studies of an {\it enhanced backscattering} \cite{WOD95}.


Let us now consider {\it many-mode} waveguides with correlated
surface disorder. In the spirit of the Landauer's concept
\cite{Land92}, the {\it total transmittance}
$\langle T\rangle$ of any quasi-1D structure is expressed as a sum
of {\it partial transmittances} $T_n$ that describe the transport
for every $n$-th propagating normal mode, $\langle
T\rangle=\sum_{n=1}^{N_d}T_n$. Here $N_d=[kd/\pi]$ is the total
number of propagating modes ({\it open channels}) determined by
the integer part $[...]$ of the ratio $kd/\pi$.

From general theory of scattering systems
\cite{BFb79} it follows that the wave propagation through
any $n$-th open channel is determined by two {\it attenuation
lengths}, the length  $L_n^{(f)}$ of forward scattering and the
backscattering length $L_n^{(b)}$. For quasi-1D waveguides with
surface disorder the inverse scattering lengths are given by,
\begin{equation}\label{Lnf}
\frac{1}{L_n^{(f)}}=\sigma^2\frac{(\pi n/d)^2}{k_nd}
\sum_{n'=1}^{N_d}\frac{(\pi n'/d)^2}{k_{n'}d}W(k_n-k_{n'}),
\end{equation}

\begin{equation}\label{Lnb}
\frac{1}{L_n^{(b)}}=\sigma^2\frac{(\pi n/d)^2}{k_nd}
\sum_{n'=1}^{N_d}\frac{(\pi n'/d)^2}{k_{n'}d}W(k_n+k_{n'}),
\end{equation}
where $k_n=\sqrt{k^2-(\pi n/d)^2}$.

The results (\ref{Lnf}) and (\ref{Lnb}) can be obtained by the
diagrammatic Green's function approach \cite{BFb79} as well as
with the use of the technique developed in Ref.~\cite{McGM84}.
Also, these expressions directly follow from the invariant
imbedding method extended to quasi-1D structures (see
Ref.~\cite{SMFY99}). Note that in a single-mode waveguide with
$N_d=1$ the sum over $n'$ contains only one term with $n'=n=1$.
Therefore, the backscattering length $L_1^{(b)}$ is four times
less then the 1D localization length $L_{loc}$, see
Eq.~(\ref{Lloc}).

The expressions (\ref{Lnf}) and (\ref{Lnb}) manifest that, in
general, both attenuation lengths are formed by scattering of a
given $n$-th propagating mode into all other modes. This is the
case when, for example, a surface profile is either
delta-correlated random function (``white noise'') with constant
power spectrum $W(k_x)$, or the function with fast decreasing
binary correlator (or, the same, with a slow decrease of its
Fourier components).

One can see that the larger the mode number $n$ is, the stronger
is the scattering of this mode into others. As was shown in
Ref.~\cite{SMFY99}, in this case a very interesting phenomenon of
the coexistence of {\it ballistic, diffusive}, and {\it localized}
transport arises, which seems to be generic for propagation
through waveguides with random surfaces. Specifically, while
lowest modes can be in the ballistic regime, the intermediate and
highest modes exhibit the diffusive and localized behavior,
respectively.

Now we are ready to show that the situation is fundamentally
different when the surface disorder has specific long-range
correlations. To demonstrate this, let us take the surface profile
$\xi(x)$ with the binary correlator in the following form, ${\cal
W}(|x-x'|)=\sin \{k_0 (x-x')\}/k_0 (x-x')$. In this case the
Fourier transform $W(k_x)$ is a "window function",
\begin{equation}\label{FTW-1}
W(k_x)=(\pi/k_0)\,\Theta(k_0-|k_x|),
\qquad k_0>0,
\end{equation}
where $\Theta(z)$ stands for the unit-step function.

In accordance with Eqs.~(\ref{beta-def}), (\ref{xi-beta}) rough
surfaces that have such correlation properties can be constructed
as follows,
\begin{equation}\label{xi-1}
\xi(x)=\frac{\sigma}{\sqrt{\pi k_0}}\,\int_{-\infty}^\infty\,dx'\,
Z(x-x')\,\frac{\sin(k_0x')}{x'}.
\end{equation}

Now one can see that in the case under consideration the number of
modes into which a given $n$-th mode is scattered, is entirely
determined by the width $k_0$ of the rectangular spectrum
(\ref{FTW-1}). It is clear that if the distance $|k_n-k_{n\pm 1}|$
between neighboring values of $k_n$ is larger than the width
$k_0$,
\begin{equation}\label{cond}
|k_n-k_{n\pm 1}|>k_0,
\end{equation}
then the transitions {\it between} all modes are forbidden. In
this case the sum over $n'$ in Eq.~(\ref{Lnf}) contains only one
term with $n'=n$ which describes scattering {\it inside} the
channels. Moreover, each term in the sum of Eq.~(\ref{Lnb}) is
equal to zero, and the scattering lengths are,
\begin{equation}\label{Lnfb-1}
\frac{1}{L_n^{(f)}}=\frac{\pi\sigma^2}{k_0}\,\frac{(\pi n/d)^4}{(k_nd)^2},
\qquad\qquad\frac{1}{L_n^{(b)}}=0.
\end{equation}

The above analysis shows that a remarkable phenomenon can arise.
Specifically, all the propagating modes with index $n$, that
satisfy the condition (\ref{cond}), are independent of other
waveguide modes, in spite of the interaction with a rough surface.
In other words, they represent a coset of {\it non-interacting}
channels. Since the transport of any 1D disordered system is
determined by the backscattering length only, and the latter is
infinite according to Eq.(\ref{Lnfb-1}), the transmittance for
such channels is $T_n=1$. This means that according to the
Landauer's formula for $\langle T \rangle$, the transmittance of
the coset of such {\it independent transparent} waveguide modes is
simply equal to their total number.

As for other propagating modes with index $n$, that do not satisfy
the condition (\ref{cond}), they remain to be mixed by surface
scattering because the roughness power spectrum (\ref{FTW-1}) is
non-zero, $W(k_n-k_{n'})=\pi/k_0$. Since these {\it mixed modes}
have finite backscattering lengths, for large enough size $L$ they
do not contribute in the total transmittance and the latter is
equal to the number of {\it independent transparent} modes.

For large number $N_d\gg1$ of open channels one can perform
further analytical analysis. In this case the inequality
(\ref{cond}) is analogous to the requirement $|\partial
k_n/\partial n|>k_0$ which can be written in the following form,
\begin{equation}\label{bal-n}
n>N_{k_0}\equiv\left[\frac{(kd/\pi)}{\sqrt{1+(k_0d/\pi)^{-2}}}\right].
\end{equation}
We remind again that square brackets stand for the integer part of
the inner expression.

The expression (\ref{bal-n}) determines the total number $N_{k_0}$
of mixed non-transparent modes, total number $N_t=N_d-N_{k_0}$ of
independent transparent modes, and the critical value of $n$ that
divides these two groups. All propagating modes with $n>N_{k_0}$
are independent and fully transparent, otherwise, they are mixed
and characterized by finite backscattering lengths $L_n^{(b)}$.

The numbers $N_{k_0}$ and $N_t$ of mixed non-transparent and
independent transparent modes are determined by two parameters,
$kd/\pi$ and $k_0d/\pi$. In the case of ``weak" correlations,
$k_0d/\pi\gg 1$, the number of mixed modes $N_{k_0}$ is of the
order of $N_d$,
\begin{equation}\label{nk0-wc}
N_{k_0}\approx\left[\left(\frac{kd}{\pi}\right)-
\frac{1}{2}\left(\frac{kd}{\pi}\right)
\left(\frac{k_0d}{\pi}\right)^{-2}\right].
\end{equation}
Consequently, in this case the number of independent modes
$N_d-N_{k_0}$ is small, or there are no such modes at all. If the
parameter $k_0d/\pi$ tends to infinity, $k_0d/\pi\to\infty$, the
rough surface profile becomes white-noise-like and, naturally,
$N_{k_0}\to N_d$.

The most interesting case is when surface profiles are strongly
correlated, which is specified by the condition $k_0d/\pi\ll 1$.
Then the characteristic integer $N_{k_0}$ is much less than the
total number of modes $N_d$,
\begin{equation}\label{nk0-sc}
N_{k_0}\approx\left[\left(\frac{kd}{\pi}\right)
\left(\frac{k_0d}{\pi}\right)\right]\ll N_d,
\end{equation}
and the number of transparent modes $N_t$ is large. When
$k_0d/\pi$ decreases and becomes small, $k_0d/\pi<(kd/\pi)^{-1}\ll
1$, the number $N_{k_0}$ vanishes and {\it all modes} become
independent and fully transparent. Evidently, if $k_0d/\pi$ tends
to zero, the roughness power spectrum $W(k_x)$, see (\ref{FTW-1}),
becomes delta-function-like and, as a consequence, $N_{k_0}=0$. In
this case the correlated disorder results in a perfect
transmission of waves, in spite of their scattering from a rough
surface.

We also would like to note that the scattering through many-mode
waveguides with correlated surfaces of the above kind reveals a
step-wise dependence of the transmittance on the value of the
parameter $\gamma_k=kd/\pi$. This is clear from the explicit
expression for the transmittance,
\begin{equation}\label{frac}
\langle T \rangle =[\gamma_k] - [\alpha \gamma_k],\,\,\,
\alpha=\{ 1+ (k_0 d/\pi)^{-2}\}^{-1/2}.
\end{equation}
The effect is similar to that known to occur for the conductance
of quasi-1D ballistic structures (see, e.g.,
\cite{vWvHB88}). Interesting enough, in our model
the step-wise dependence of the transmittance arises both for
integer and {\it non-integer} values of $\gamma_k$, due to the
presence of the second term in Eq.(\ref{frac}).

In conclusion, we have studied transport properties of quasi-1D
waveguides with random correlated surfaces. It was analytically
shown that for a specific choice of long-range correlations a
subset of propagating modes emerges that are {\it independent} and
fully {\it transparent}. These modes do not interact with each
other and the theory of 1D transport can be rigorously applied for
them. The rest of propagating modes creates the second set of {\it
mixed} and {\it non-transparent} modes. With a proper choice of
model parameters, one can construct such waveguides for which {\it
all} modes turn out to be completely delocalized. This means a
perfect transparency of waveguides in spite of random character of
surface scattering. These results may be used for experimental
fabrication of waveguides and supperlattices with selective
windows of a complete transparancy in dependence of the wave
number of incoming waves.

This research was supported by Consejo Nacional de Ciencia y
Tecnolog\'{\i}a (CONACYT, M\'exico) grant 34668-E, and by the
Universidad Aut\'onoma de Puebla (BUAP, M\'exico) under the grant
II-63G01.

\end{document}